\documentclass[a4paper]{article}

\usepackage{ISCSLP2021}
\usepackage{multirow}
\usepackage{cite}
\usepackage{url}

\title{Adversarial Training for Multi-domain Speaker Recognition}
\name{Qing Wang$^1$, Wei Rao$^2$, Pengcheng Guo$^1$, Lei Xie$^{1, *}$\thanks{*Lei Xie is the corresponding author.}}
\address{
  $^1$Audio, Speech and Language Processing Group (ASLP@NPU), \\
  School of Computer Science, Northwestern Polytechnical University, Xi'an, China\\
  $^2$Tencent Media Lab, Shenzhen, China}
\email{\{qingwang, pcguo, lxie\}@nwpu-aslp.org, ellenwrao@tencent.com}

\begin{document}

\maketitle
\begin{abstract}
  In real-life applications, the performance of speaker recognition systems always degrades when there is a mismatch between training and evaluation data. Many domain adaptation methods have been successfully used for eliminating the domain mismatches in speaker recognition. However, usually both training and evaluation data themselves can be composed of several subsets. These inner variances of each dataset can also be considered as different domains. Different distributed subsets in source or target domain dataset can also cause multi-domain mismatches, which are influential to speaker recognition performance.
  In this study, we propose to use adversarial training for multi-domain speaker recognition to solve the domain mismatch and the dataset variance problems.
  %, which makes the speech representations more efficient. 
  By adopting the proposed method, we are able to obtain both multi-domain-invariant and speaker-discriminative speech representations for speaker recognition.
  %To cope with these problems, we learn the multi-domain-invariant and speaker-discriminative speech representations via adversarial training method. 
  Experimental results on DAC13 dataset indicate that the proposed method is not only effective to solve the multi-domain mismatch problem, but also outperforms the compared unsupervised domain adaptation methods.
  
  %We get 36\% and 6\% relative improvement of EER on 2013 domain adaptation challenge (DAC13) dataset, compared with no domain adaptation baseline and our previous work (DAT), respectively.
  %Moreover, this study compares the proposed method with other state-of-the-art unsupervised domain adaptation techniques for i-vector based speaker recognition. Experiments demonstrate that 
\end{abstract}

\section{Introduction}
% In recent years, speaker recognition is becoming an important topic for bio-metric, and the performance of speaker recognition system is consistent with i-vectors~\cite{Dehak2011Front} and x-vectors~\cite{snyder2018x} when PLDA~\cite{Prince2007Probabilistic} is used as a score function.
In recent years, speaker recognition is becoming an important topic for biometrics, and a lot of techniques have been studied to improve its performance, such as i-vectors~\cite{Dehak2011Front} and x-vectors~\cite{snyder2018x}.
However, like many machine learning tasks, speaker recognition system can be susceptible to performance degradation when substantial mismatch exists between the training and evaluation data.
%when the training and evaluation data are from different distributions, the performance of speaker recognition will degrade dramatically. Domain adaptation is one of the branches of transfer learning.
The population of interest (the evaluation data), is called target domain, for which the labels are usually not available, so it's impossible to train a classifier directly. At the same time, if data from anther population (the training data), whose labels are available, it could be used as source domain data. Thus, \textit{domain adaptation}~\cite{Kouw2018an} is a common method to overcome the differences between domains so that the classifier trained on the source domain data generalizes well to the target domain data. When target domain labels are not available for training or fine-tuning, unsupervised domain adaptation becomes an alternative to improve the source domain trained model. 

In speaker recognition area, many domain adaptation approaches have been successfully used to solve the domain mismatch problem~\cite{mclaren2011source,glembek2014domain,singer2015domain,Shon2017Autoencoder}.
Most of these domain adaptation methods assume that the examples inside source or target domains are from the same distributions, however, usually the training and evaluation data are composed of several different datasets. Considering these data as a single domain will cause a sub-optimal solution, so eliminating mismatches among the multiple domain is vital. 
Multiple domain adaptation aims at obtaining a model with minimal average risk across multiple domains. Several studies have addressed this problem. In inter-dataset variability compensation (IDVC)~\cite{Aronowitz2014Inter}, Aronowitz et al. found that the source domain dataset and target domain dataset could be composed of several subsets and there would be variances between the subsets. In~\cite{pei2018multi}, Pei et al. presented a multi-adversarial domain adaptation approach, which captures multimode structures to enable fine-grained alignment of different data distributions based on multiple domain discriminators instead of just based on the single discriminator. The adaptation can be achieved by stochastic gradient descent with the gradients computed by back-propagation in linear-time. 
In~\cite{lin2018multisource}, Lin et al. proposed to use maximum mean discrepancy (MMD) to solve the problem of multiple source domains in speaker recognition. They adopted MMD to measure the discrepancies among multiple distributions and incorporated the discrepancy term into the objective function for training auto-encoders.

Domain adversarial training (DAT) was proposed by Ganin et al.~\cite{Ganin2014Unsupervised,ganin2016domain} to eliminate the domain mismatch in computer vision task, which inserted a gradient reverse layer (GRL) to learn both class-discriminative and domian-invariant intermediate representations in a multi-task framework.
%r unsupervised domain adaptation because of its great success in estimating generative models [14]. In typical DAT approaches [15, 16, 17, 18], the deep neural network (DNN) is transformed to three sub-networks with different learning purposes: a classifier C (predicting senones) works in parallel with a domain discriminator D (predicting the input whether from source or target domain) and a feature generator G is shared at the bottom. G and C are trained to minimize the class prediction error and maximize the domain prediction error respectively, while D is updated to minimize the domain prediction error. In a form of multi-task learning, a gradient reverse layer (GRL) is inserted to learn both class-discriminative and domian-invariant intermediate representations. 
In recent studies, adversarial learning has been widely applied in speech related tasks~\cite{houna2019domain,wang2019adversarial,wang2020inaudible,xia2019cross}. In speech recognition, Sun et al.~\cite{sun2018domain} proposed to use DAT in accented speech recognition. In~\cite{guo2019unsupervised}, Guo et al. further refined the common DAT method by introducing an adversarial dropout regularization term. Hou et al.~\cite{hou2019domain} applied DAT to improve keyword spotting performance of ESL speech. In speaker recognition area, Wang et al.~\cite{wang2018unsupervised} firstly proposed to use DAT to project source domain data and target domain data into a common space and extract domain-invariant and speaker-discriminative speech representations. %Xia et al.~\cite{xia2019cross} introduced unsupervised adversarial discriminative domain adaptation to solve the language mismatch. 
Fang et al.~\cite{fang2019channel} learned the channel-invariant and speaker-discriminative representations via adversarial training. In~\cite{tu2019variational,tu2020variational}, Tu et al. proposed variational domain adversarial neural network and information-maximized variational domain adversarial neural network to reduce domain mismatch. 

%domain adaptation to project source domain data and target domain data into a common space in speaker recognition. Gradient reversal layer is used to confuse the main network so that domain-invariant and speaker-discriminative speech representations can be extracted to improve the performance of the speaker recognition system.

In this study, we propose to use adversarial training for multi-domain speaker recognition. 
Due to the source and target domain data sets can be decomposed into several different subsets, instead of just solving the mismatch problem between source and target sets, we also take the inner variances of each dataset into consideration. %by decomposing them into several different subsets.
% , which the training and evaluation set might be composed by several different subset. 
%不知道这句话怎么加了，就是如果把source 或者target domain各自算成自己的domain 会导致sub-optimal：In many application scenarios, the source data may come from multiple domains with different distributions. Considering these data as single source will cause sub-optimal solution.
In other words, we regard each subset as an independent domain and address this problem as a multi-domain mismatch in speaker recognition. Our goal is to reduce the differences among all the domains. 
In this multi-domain speaker recognition task, we mainly focus on three conditions of domain adaptation: multiple source domains with one target domain; one source domain with multiple target domains; as well as multiple source domains with multiple target domains. In order to extract multi-domain-invariant and speaker-discriminative speech representations, we use adversarial training for multi-domain adaptation. During training, we use a multi-task framework, while the first task is the speaker classifier and the second task is the multi-domain discriminator. With gradient reversal layer (GRL) added before the domain discriminator, the main network is not sensitive to multiple domains and projects the data from different domains into a same subspace. Experiments on 2013 domain adaptation challenge (DAC13) demonstrate that our proposed method can eliminate the mismatches among all the domains.

\section{Proposed method}

\subsection{Multi-domain adaptation by adversarial training}

In unsupervised multi-domain adaptation task, we only have the access to a source domain example $\mathbf{x}_{s_n}$ and its corresponding label $y_{s_n}$ drawn from different labeled sub-source domain sets $\{\mathbf{X_s}, \mathbf{Y_s}\} = \{(\mathbf{X}_{s_1}, \mathbf{Y}_{s_1}),...,(\mathbf{X}_{s_N}, \mathbf{Y}_{s_N})\}$, in which $N$ means the number of sub-source domains. And a target samples $\mathbf{x}_{t_m}$ drawn from different unlabeled sub-target domain sets 
$\{\mathbf{X_t}\} = \{\mathbf{X}_{t_1},...,\mathbf{X}_{t_M}\}$, in which $M$ means the number of sub-target domains. The source domain and target domain set themselves are decomposed into several subsets, which have variances among each others. 
%The distributions between source and target domains are different, while there are mismatches among the subset of the source or target domain. 
The mismatch problem is not only between the distribution of source and target domains, but also among the subsets from the source and target domains. In order to solve these mismatch problems, we try to eliminate the mismatch between the source and target domain as well as to reduce the variance among different subsets. 
% And we call it \textit{multiple domain adaptation}.

%\subsection{Multiple domain adaptation by adversarial training}

% In unsupervised multiple domain adaptation for speaker recognition, we 
Assume that we have the access to a set of labeled source domain sample comes from $N$ sub-source domains and a set of unlabelled target domain samples from $M$ sub-target domains as defined before. When the number of source domain $N$ and target domain $M$ are both equal to 1, the framework is the same as conventional DAT in~\cite{wang2018unsupervised}. In order to eliminate the multi-domain mismatches and dataset variances, we project the multiple domains into a common subspace by multi-domain adversarial training, which aims to learn speaker-discriminative and multi-domain-invariant feature representations. The details of the proposed method are shown in the following context.

\begin{figure}[htb]
\begin{minipage}[b]{1.0\linewidth}
  \centering
  \centerline{\includegraphics[width=8.8cm, height=7.05cm]{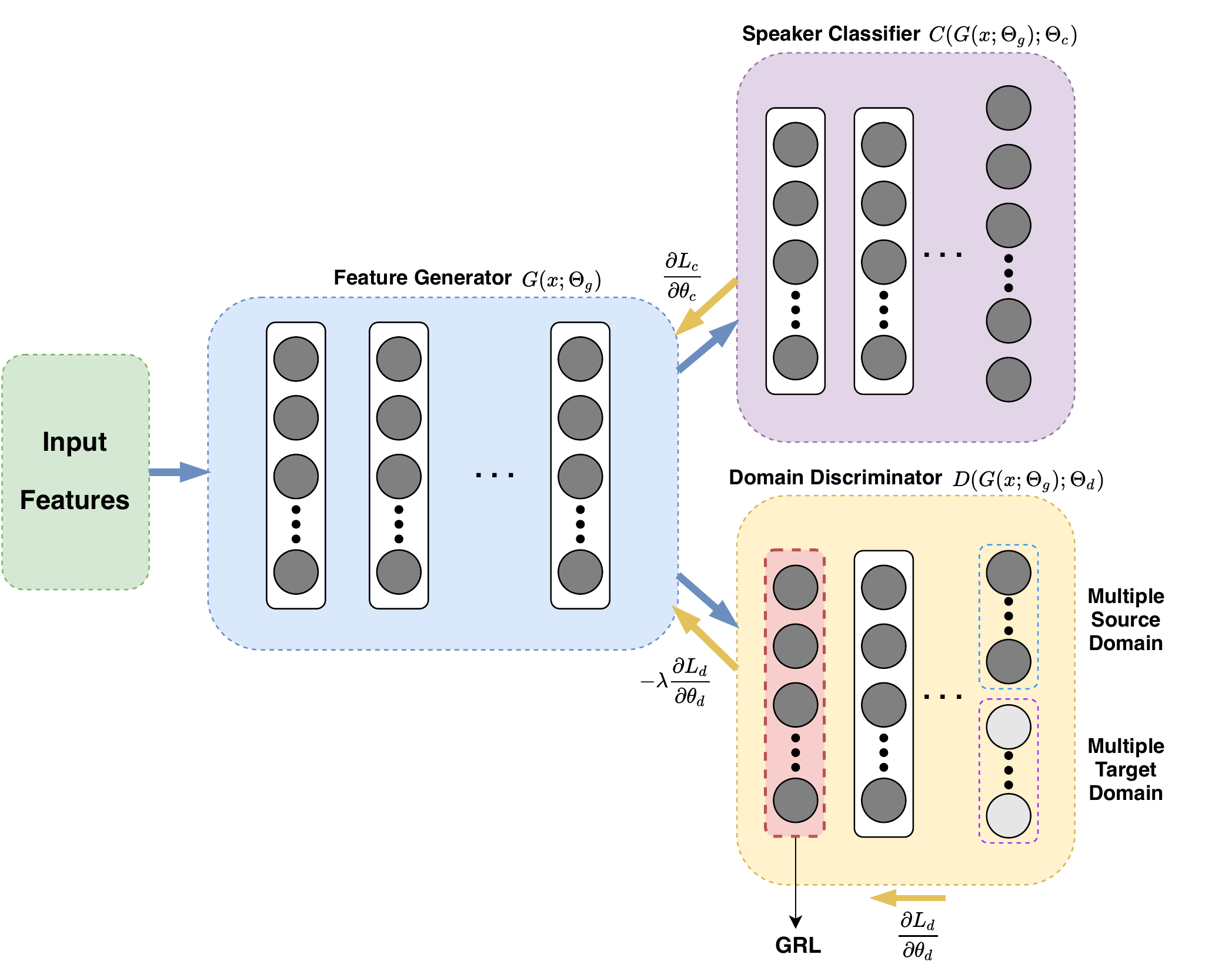}}
\end{minipage}
\caption{An overview of adversarial training for multi-domain speaker recognition framework.}
\label{fig:res}
\end{figure}

%When the number of source domain N and target domain M are both equal to 1, the framework is the conventional domain adversarial training network in our previous work~\cite{wang2018unsupervised}.

% In multiple domain adaptation for speaker recognition, we are given source domain data with speaker labels and target domain data without speaker label, as well as the number of sub-source domains and the number of sub-target domains (if we have the prior knowledge of the target domain). In order to eliminate the domain mismatch and dataset variance, we project the multiple domains and subsets into a common subspace by using the domain adversarial training (DAT) method which learns a speaker-discriminative and multiple-domain-invariant feature representation. The details of Multi-DAT are shown in the following context.

%We assume that we have the access to a set of labeled source domain sample comes from $N$ sub-source domains and unlabelled target domain samples from $M$ sub-target domains as we defined before. When the number of source domain N and target domain M are both equal to 1, the framework is the conventional domain adversarial training network in our previous work~\cite{wang2018unsupervised}.

The model can be decomposed into three parts as shown in Figure~\ref{fig:res}: a feature generation network $G$, which takes $\mathbf{x}_{s_n}$ or $\mathbf{x}_{t_m}$ as the input; a speaker classifier $C$, which takes source domain samples' features from $G$ and classifies them into $K$ speaker classes; as well as a domain discriminator $D$, which predicts the specific domain label of the feature from $G$. 
%In supervised multiple domain adaptation, we have the access of the label of target domain data, and we used to train the speaker classification network as well. However, in
For unsupervised domain adaptation task, only the labeled source domain data are used to train the $C$ and both source domain and target domain data are used to update the $D$. The mapping functions can be formulated as: $G(\textbf{x};\Theta_g)$, $C(G(\textbf{x};\Theta_g);\Theta_c)$ and $D(G(\textbf{x};\Theta_g);\Theta_d)$, where $\Theta_g$, $\Theta_c$ and $\Theta_d$ are the parameters of the networks.

The following two requirements are satisfied in our method. First, $C$ and $D$ are estimated to ensure that they will perform accurate speaker and domain classifications. Second, $G$ is used to extract speaker-discriminative and multi-domain-invariant speech representations.

The objective function of the proposed model consists of two parts. In the first part, $G$ and $C$ have to classify source samples into $K$ classes of speakers correctly to obtain the speaker-discriminative features. Thus, we update both networks' parameters based on the following standard classification loss. Given a source domain sample $\mathbf{x}_{s_n}$ and its speaker labels $y_{s_n}$, the objective function is:

\begin{equation}
    \min_{G, C} L_{cls} = - \mathbb{E}_{(\mathbf{x}_{s_n}, y_{s_n}) \sim (\mathbf{X}_s, \mathbf{Y}_s)} \sum^{K}_{k=1} y_k \log C(G(\mathbf{x}_{s_n}))_k,
\end{equation}
where $C(G(\textbf{x}_{s_n}))_{k}$ returns the probability that the sample $\textbf{x}_s$ is assigned to the class $k$.

In the second part, $D$ is trained as a discriminator to predict the domain label of all the samples. The objective function of this part is: 

\begin{equation}
         \min_{D} L_{adv} = - \mathbb{E}_{\textbf{x} \sim \{\textbf{X}_s, \textbf{X}_t\}} \sum^{N+M}_{i} d_i \log D(G(\textbf{x}))_{i},
\end{equation}
where $d_i$ refers to the domain label and $D(G(\textbf{x}))_{i}$ returns the probability of the sample $\textbf{x}$ belonging to the $i$-th domain.

We are supposed to jointly train the $G$, $C$ and $D$, and we need to seek a $\Theta_g$ to minimize the speaker classification's objective function and to maximize the domain discriminator's objective function at the same time. Since $G$ should learn the speaker-discriminative features for source domain samples as well as multi-domain-invariant features for the samples from all the domains. There is a gradient reversal layer (GRL)~\cite{ganin2016domain} added between the feature generator $G$ and the domain discriminator $D$ to search a saddle point between $C$ and $D$. The GRL is multiplied by a certain $\lambda$ during the back-propagation. $\lambda$ is a positive hyper-parameter used to trade off the losses. Gradient reversal layer ensures the feature distributions over different domains approaching similar so that we can get multi-domain-invariant and speaker-discriminative speech embeddings.
The final objective function is defined as: 

\begin{equation}
    \min_{G, C} \max_{D} L = L_{cls} - \lambda L_{adv}.
\end{equation}

% \begin{equation}
% \begin{split}
% \min_{G, C} \max_{D} L(X, Y) =&- \mathbb{E}_{(\textbf{x}_{s_n}, \textbf{y}_s) \sim (\textbf{X}_{s_n}, \textbf{Y}_s)} \sum^{K}_{k=1}\log C(G(\textbf{x}_{s_n}))_{k}\\
% &+ \lambda \mathbb{E}_{(\textbf{x}, \textbf{y}) \sim (\textbf{X}, \textbf{Y})} \sum^{N+M}_{i=1}\log D(G(\textbf{x}))_{n+m}
% \end{split}
% \end{equation}
%\begin{equation}
%    \begin{align}
%         \min \limit_{G, C} \max \limit_{D} L(X, Y) = - \mathbb{E}_{(\textbf{x}_{sn}, \textbf{y}_s) \sim (\textbf{X}_{sn}, \textbf{Y}_s)} \sum^{K}_{k=1}\log C(G(\textbf{x}_{sn}))_{k} \\
%         + \lambda \mathbb{E}_{(\textbf{x}, \textbf{y}) \sim (\textbf{X}, \textbf{Y})} \sum^{N+M}_{i=1}\log D(G(\textbf{x}))_{n+m}
%    \end{align}
%\end{equation}
%\begin{equation}
%    \limit L(X, Y) = L(X_s, Y_x) - L(X, Y)
%\end{equation}
We optimize with back-propagation using following formulas:

\begin{equation}
    \theta_g \longleftarrow \theta_g - \mu(\frac{\partial L_{cls}}{\partial \theta_g} - \lambda \frac{\partial L_{adv}}{\partial \theta_g}),
\end{equation}
\begin{equation}
    \theta_c \longleftarrow \theta_c - \mu(\frac{\partial L_{cls}}{\partial \theta_c}),
\end{equation}
\begin{equation}
    \theta_d \longleftarrow \theta_d - \mu(\frac{\partial L_{adv}}{\partial \theta_d}),
\end{equation}
where $\mu$ is the learning rate.

%\subsection{Supervised multiple adversarial domain adaptation}

%If the speaker label of target domain is available, 

%\subsection{Dynamic $\lambda$}

%When we optimize the hyper-parameter $\lambda$, in order to suppress noisy signal from the domain discriminator at the early stages of the training procedure instead of fixing the adaptation factor $\lambda$~\cite{ganin2016domain}, we gradually change it from 0 to 1 using the following schedule:

%\begin{equation}
%   \lambda_p = \frac{2}{1+exp(-\gamma \cdot p)}-1 
%\end{equation}
%where $\gamma$ was set to 10 in all experiments (the schedule was not optimized/tweaked).

\subsection{Adversarial training for multi-domain speaker recognition}

%\subsection{Extracting  Speaker-Discriminative  and  Multi-Domain-Invariant Speech Representations}
%We usually follow this strategy to perform the speaker recognition. feature extraction, 
%For speaker recognition, if we have the access of the target domain labels, we use the supervised multi-DAT, otherwise, we adopt unsupervised multi-DAT. The input of the multi-DAT model can be any kinds of features, such as MFCCs, i-vectors~\cite{Dehak2011Front} or x-vectors~\cite{snyder2018x}. The feature generator is used to generate common space new speech embeddings. 

In speaker recognition, the data we used to train the PLDA~\cite{Prince2007Probabilistic} are usually supposed to share the same distribution with the evaluation data, which are defined as target domain. However, in many scenarios, the target domain data are insufficient or the speaker labels are unavailable. So we propose to project multiple source and target domains into a common subspace, and we can use the projected source domain data with speaker labels for the speaker recognition.

\begin{figure}[t]
  \centering
  \includegraphics[width=8cm]{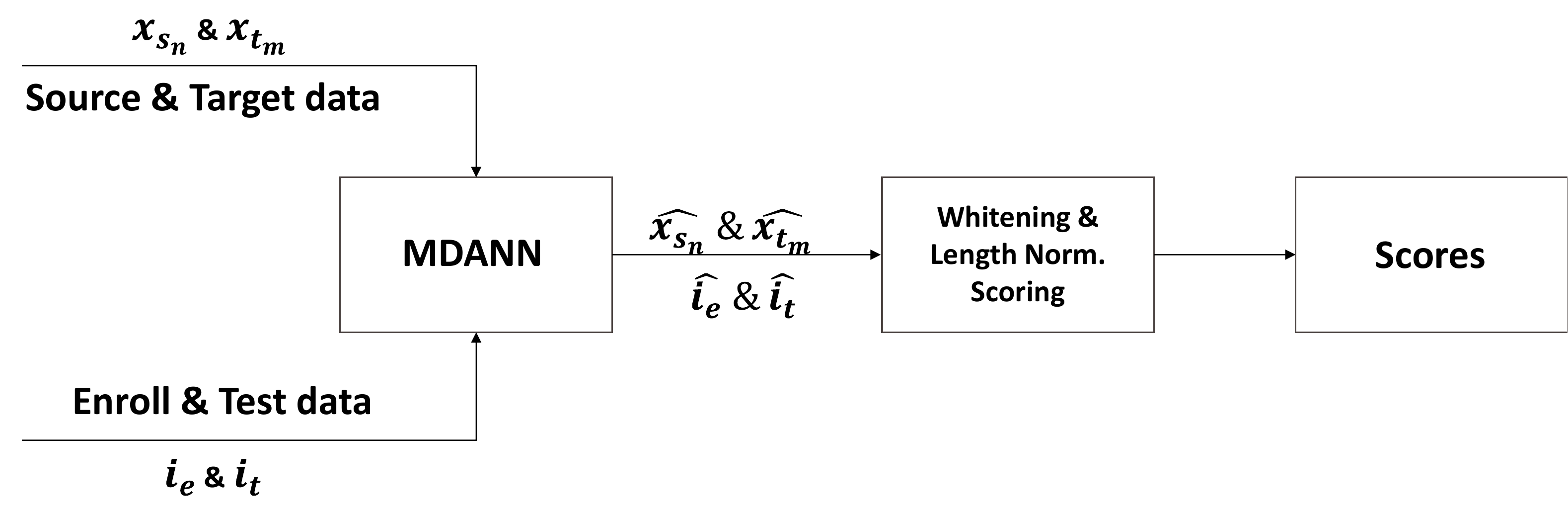}
  \caption{Block diagram of adversarial training for multi-domain speaker recognition.}
  \label{fig:MAUDA-score}
  %\vspace{-12pt}
\end{figure}

First, we use multiple source and target data to train the multi-domain adaptation neural network (MDANN). Since we don't have speaker label of the target domain data, only the multi-source domain data is used to train the first task. The label of the first task is speaker ID. %For the second task training, we use both of source and target domain data.
All the multi-source and multi-target domain data are used to train the second task. The label of the second task is subsets labels. 
%With GRL, we can obtain
After training the MDANN, we use multi-source domain vectors ($\textbf{x}_{s_n}$), multi-target domain vectors ($\textbf{x}_{t_m}$), enroll vectors ($\textbf{i}_e$) and test vectors ($\textbf{i}_t$) as the inputs to the MDANN and we use the representation of the hidden layer of the feature generation network as the new vectors ($\hat{\textbf{x}}_{s_n}$, $\hat{\textbf{x}}_{t_m}$, $\hat{\textbf{i}}_{e}$ and $\hat{\textbf{i}}_{t}$) of all the data. The extracted embeddings are therefore expected to be multi-domain-invariant and speaker-discriminative speech representations which stand in the same subspace. Then, we apply the pre-processing (whitening and length-normalization~\cite{Garcia2011Analysis}) to the $\hat{\textbf{x}}_{s_n}$, $\hat{\textbf{x}}_{t_m}$, $\hat{\textbf{i}}_{e}$ and $\hat{\textbf{i}}_{t}$. Finally, we use a scoring function to compute the scores between the speaker model and the test sample. In this paper, we adopt PLDA as the scoring method. Figure~\ref{fig:MAUDA-score} shows how we use the proposed strategy in speaker recognition. 

\section{Experimental setup}

\subsection{Dataset}

We use the 2013 domain adaptation challenge (DAC2013) dataset~\cite{JHU2013SpeakerRecognitionWorkshop} as the evaluation data set.
%In this paper, we use 2013 domain adaptation challenge dataset, donated as DAC 2013~\cite{JHU2013SpeakerRecognitionWorkshop} as the unsupervised multiple domain adaptation task evaluation dataset and VOICES dataset~\cite{} as the supervised multiple domain adaptation task evaluation dataset.
%We use DAC 2013 i-vector dataset as the unsupervised training and evaluation dataset. 
%In DAC 2013, they posed a domain adaptation task based on LDC telephone corpora which demonstrates
DAC2013 is designed based on LDC telephone corpora and targeted the domain adaptation. It focuses on the effect of dataset mismatch on hyper-parameters, such as the latent speaker and channel factors for PLDA. Both the audio lists and i-vectors of NIST SRE data~\cite{martin2010nist} and Switchboard data~\cite{godfrey1992switchboard} are provided. For fair comparison with other domain adaptation techniques, DAC2013 i-vector dataset is used in this study. Table~\ref{table:dataset} shows the details of DAC2013 i-vector dataset.
%which was provided by MITLL~\cite{}. 
The dimension of i-vectors is 600. Two datasets are defined for hyper-parameter training: 1) the source domain SWB set consists of all telephone calls from all speakers taken from the Switchboard-\uppercase\expandafter{\romannumeral1} and Switchboard-\uppercase\expandafter{\romannumeral2} (all phases) corpora; 2) the target domain SRE set consists of all the telephone calls without speaker labels are taken from the NIST SRE 04, 05, 06, and 08 collections, while SRE-1phn is a reduced set of SRE with only the i-vectors from 1 telephone number per speaker, which makes it hard to estimate within-class variability because of the lack of speaker and channel information. In this study, we selected the more challenging SRE-1phn data for the domain adaptation task. The telephone data of NIST SRE 2010 (SRE10) was selected as the evaluation set of DAC2013.

\vspace{-8pt}
\begin{table}
\centering
\caption{i-vector statistic in DAC 13 i-vector dataset}
\vspace{-8pt}
\label{table:dataset}
\begin{tabular}{cccc} 
\toprule
                            & \textbf{SWB} & \textbf{SRE} & \textbf{SRE-1phn}  \\ 
\hline\hline
\textbf{\#spks}             & 3114         & 3790         & 3787               \\
\textbf{\#calls}            & 33039        & 36470        & 25640              \\
\textbf{\#calls/spkrs}      & 10.6         & 9.6          & 6.77               \\
\textbf{\#phone\_num/spkrs} & 3.8          & 2.8          & 1.0                \\
\bottomrule
\end{tabular}
\vspace{-12pt}
\end{table}
%\vspace{-8pt}
%Table 1 shows the details of DAC2013 i-vector dataset. In the original DAC2013 i-vector dataset, it did not provide the development set. To tune the network, this paper randomly selected 50 speakers from SRE as the development set.

%And we randomly choose 50 speakers in target domain dataset as the development set to optimize the network.

%%\begin{table}%{h!}%{0.5\textwidth}
%%\caption{i-vector Statistic in DAC 13 i-vector Dataset}
%%\vspace{0.1cm}
%%\label{tab:1} 
%%\centering
%%\begin{tabular}{|p{0.1\textwidth}|p{0.1\textwidth}|p{0.1\textwidth}|p{0.1\textwidth}|} 
%% \hline
%%  & SWB & SRE & SRE-1phn \\[0.5ex]
%% \hline
%% \#spks & 3114 & 3790 & 3787 \\[0.5ex]
%% \hline
%% \#calls  & 33039 & 36470 & 25640 \\[0.5ex]
%% \hline
%% \#calls/spkrs & 10.6 & 9.6 & 6.77 \\[0.5ex]
%% \hline
%%\#phone\_num/spkrs & 3.8 & 2.8 & 1.0 \\[0.5ex]
%% \hline
%%\end{tabular}
%%\end{table}

\subsection{Data partition for multi-domain}
%In many application scenarios, the source data may come from multiple domains with different distributions. Considering these data as single source will cause sub-optimal solution. This paper propose MS-DAT to solve this problem. 
Data partition is performed in two kind of manners. Firstly, following the work in~\cite{Aronowitz2014Inter}, we divide the source domain data into 6 subsets based on the different LDC codes (LDC97S62 Switchboard-1 Release2, LDC98S75 Switchboard-2 Phase~\uppercase\expandafter{\romannumeral1}, LDC99S79 Switchboard-2 Phase~\uppercase\expandafter{\romannumeral2}, LDC2002S06 Switchboard-2 Phase~\uppercase\expandafter{\romannumeral3} Audio, LDC2001S13 Switchboard Cellular Part 1 Audio and LDC2004S07 Switchboard Cellular Part2 Audio). 
The second data parition manner is clustering. Because some of the data are recorded in the same year, but released in different subsets. We use k-means to cluster the source domain data into 3 subsets. %The second strategy we use to divide the data is using the k-means to cluster the source domain data into 3 subsets.

%We adopt the same strategies as mentioned above when we partial the target domain data. Because the target domain is unlabeled, when we have the prior information of the target domain data's LDC code (SRE2004, SRE2005, SRE2006, SRE2008). The target domain subsets will be partitioned based on the different LDC distributions. 
Because the target domain data is usually unlabeld, multi-target domain labels may not be accessed. So if we don't have the prior information of the target domain, we did not partial the target domain data. 
When the prior information is available, we adopt the same strategies as mentioned above to divide the target domain data. The first one is based on the subset's LDC codes (SRE2004, SRE2005, SRE2006, SRE2008). As a result, the target domain data is divided into 4 subsets according to the LDC distributions.
Furthermore, k-means is also used to cluster the target domain data into 2 subsets.

\subsection{Experimental setup}

In unsupervised adversarial training for multi-domain speaker recognition experiments, the way we define the classes of multi-source and multi-target domain is shown in Table~\ref{table:1}, where $N$ and $M$ indicate the numbers of multi-source domains and multi-target domains respectively. %The conditions that only the source or the target domains are distinguished during adaptation, are denoted as MS-DAT and MT-DAT, respectively. 
The conditions that with the multiple source domains only or with the multiple target domains only are denoted as MS-DAT and MT-DAT, respectively. 
In addition, MDAT is the condition of both source and target domains have multiple subsets.

%\vspace{-8pt}
\begin{table}
\centering
\caption{Unsupervised multiple domain adaptation with adversarial training setup}
\vspace{-8pt}
\label{table:1}
\begin{tabular}{ccc} 
\toprule
\textbf{~Adaptation Methods~}              & \textbf{~ ~ $N$~ ~~} & \textbf{~ ~ $M$~ ~~}  \\ 
\hline\hline
DAT~\cite{wang2018unsupervised}                            & 1                  & 1                   \\ 
\hline
~ MS-DAT (based on LDC code)~~ & 6                  & 1                   \\
MT-DAT (based on LDC code)     & 1                  & 4                   \\
MDAT (based on LDC code)       & 6                  & 4                   \\
MS-DAT (based on k-means)      & 3                  & 1                   \\
MT-DAT (based on k-means)      & 1                  & 2                   \\
MDAT (based on k-means)        & 3                  & 2                   \\
\bottomrule
\end{tabular}
\vspace{-12pt}
\end{table}

For fair comparison, we use the same structure to all the DAT systems and we follow the setup in~\cite{wang2018unsupervised}. In the multi-domain adversarial neural network, there are two fully connected layers with 512 nodes for feature generation network $G$, two fully connected layers with 300 hidden nodes for speaker classification network $C$, and two fully connected layers with 512 hidden nodes for domain discrimination network $D$. %The dynamic $\lambda$ is used in our experiments and the effect of different values of $\lambda$ will shown in Section 4.4.

In the training stage, we use SWB with speaker labels as multi-source domain data and SRE-1phn with no speaker label as multi-target domain data for the MDANN training, and use the subset domain labels as the supervised information of the second task. 
During the test stage, we use SWB, SRE-1phn, enroll and test data as the inputs of the network and extract the multi-domain-invariant and speaker-discriminative speech representation from the first hidden layer of the feature generation network $G$. After that the new SRE-1phn data is used for the whitening and centering to all the new data, and the PLDA back-end is trained using new SWB data with its speaker labels to obtain the scores.

\section{Experimental results and analysis}

\subsection{Result of baseline}
Table~\ref{tab:baseline} shows the performance of the SRE10 c2-extended test when the parameters are trained with different datasets using the i-vector PLDA framework. System 1 can be considered as the desired benchmark when the in-domain dataset speaker label is known. System 2 is the baseline of the domain mismatched condition when the in-domain database is unlabeled. Systems 3 and 4 are versions of systems 1 and 2, respectively on more challenging conditions. System 3 is adapted using a matched in-domain labeled dataset SRE-1phn which is subset of SRE. Note that, although system 3 is under domain matched conditions and system 4 is under mismatched conditions, system 4 shows better performance in EER than system 3.
This is an interesting result and we believe that performance was degraded by insufficient channel information. The dataset ‘SRE-1phn’ contains audio from only a single telephone number per speaker and use of such a poor phone number
diversity hinders the effective estimation of within-speaker variability of in-domain. %In this case, the conventional approaches [13], [14] that estimate within-speaker variability from in-domain unlabeled dataset would fail, in spite of perfect speaker label estimation, due to insufficient channel information.

% \begin{table}%{h!}%{0.5\textwidth}
% \caption{Results of DAC 2013 i-vector dataset without domain adaptation}
% \vspace{0.1cm}
% \label{tab:1} 
% \centering
% \begin{tabular}{|p{0.07\textwidth}|p{0.05\textwidth}|p{0.05\textwidth}|p{0.05\textwidth}|p{0.05\textwidth}|p{0.05\textwidth}|} 
%  \hline
%  Systems\# & Pre-processing & PLDA & EER\% & DCF10 & DCF08   \\
 
%  \hline
%  1 & SRE & SRE & 2.33 & 0.402 & 0.235\\[0.5ex]
%  \hline
%  2 & SRE & SWB & 5.65 & 0.632 & 0.427\\[0.5ex]
%  \hline
%  3 & SRE-1phn & SRE-1phn & 9.35 & 0.724 & 0.520\\[0.5ex]
%  \hline
%  4 & SRE-1phn & SWB & 5.66 & 0.633 & 0.427\\[0.5ex]
%  \hline

% \end{tabular}
% \end{table}
%\vspace{-8pt}
\begin{table}[h]
	\caption{Results of DAC 2013 i-vector dataset without domain adaptation}
	\vspace{-8pt}
	\label{tab:baseline}
	\centering
	\begin{tabular}{ccccc}
		\toprule
		\textbf{Systems\#} & \textbf{PLDA}  & \textbf{EER (\%)} & \textbf{DCF10}   & \textbf{DCF08}       \\
		\hline\hline
		1 & SRE & 2.33 & 0.402 & 0.235\\
		%\hline
		2 & SWB & 5.65 & 0.632 & 0.427\\
		%\hline
		3 & SRE-1phn & 9.35 & 0.724 & 0.520\\
		%\hline
		4 & SWB & 5.66 & 0.633 & 0.427\\
		\bottomrule
	\end{tabular}
	\vspace{-12pt}
\end{table}
%\vspace{-8pt}
%\subsection{Baseline experimental result using Domain adversarial training}
%Our another baseline is using domain adversarial training for speaker recognition proposed in our previous work~\cite{wang2018unsupervised}. we consider SWB as the source domain data with speaker lables and SRE-1phn data as the target domain data without speaker label. In baseline system, i-vector pre-processing are done by SRE-1phn first. Training data of DANN consists of two parts: SWB i-vectors with speaker labels and SRE-1phn i-vectors without speaker label. SWB data are used to train the whole network while the SRE-1phn i-vectors are used to train the feature extractor and the domain classifier. Because the data from target domain lack of speaker labels, we randomly generate speaker labels for the target domain data in order to train the model in a uniform framework. And we use a binary flag~\cite{Sun2017An} to control the target domain data do not train in the speaker label predictor.  At the test stage, we use SWB, SRE-1phn, enroll and test data to forward the network and extract the first hidden layer's representation as the new speech representation which has been projected into the same subspace. After that we do the same back-end to obtain the scores.

\subsection{Result of proposed method}

The experimental results are given in Table~\ref{table:mdat}. As a comparison, the first five systems are other domain adaptation methods~\cite{Shon2017Autoencoder} in speaker recognition.
The system DAT is when the number of sub-source and sub-target domain are both equal to 1, which is the conventional domain adversarial neural network in~\cite{wang2018unsupervised}.
The last six systems are multi-domain adversarial training experiments. The configurations of them are shown in Table~\ref{table:1}.
The experimental results demonstrate that, by projecting the multi-source domain data and multi-target domain data to a common space with the proposed approach, we can achieve the lowest EER, DCF10 and DCF08 compared to the conventional DAT~\cite{wang2018unsupervised} and other adaptation methods~\cite{Shon2017Autoencoder}. And almost all the systems of our method outperform the conventional DAT.
%The multi-domain configurations of the last seven systems are shown in Table~\ref{table:1}. The system DAT is when the number of sub-source and sub-target domain are both equal to 1, which is the domain adversarial neural network in~\cite{wang2018unsupervised}.
%For a comparative study, we also report the results of first five adaptation methods for speaker recognition in~\cite{Shon2017Autoencoder}.  
%By projecting the multiple source domain data and target domain data to a common space with the proposed approach, our method almost outperforms the conventional DAT in EER, DCF10 and DCF08.
%For  System  System 11 and System 14, multiple source domains and one target domain.  
%We observe that, by projecting the multiple source domain data and target domain data to a common space with MSUDA approach, MSUDA (System 11) in Table~\ref{tab:1} shows a 36\% improvement over the System 4 baseline on EER and a 3\% improvement over our previous work DANN (System 10). And after we project the multiple source and target domain data into a common space using MAUDA approach (System 12), we can get a 33\% improvement over the System 4 baseline system on EER, and a 10.4\% improvement over our previous DANN system in DCF08.
%The experimental results of single source domain and multiple target domains (System 12 and System 15)

When the inner variances of both source and target domain are considered, we can get the best performance in system MS-DAT (based on LDC code or k-means). With the adversarial training for multi-domain speaker recognition, the EER is improved from 5.66\% to 3.58\%, with +36.7\% relative error reduction compared to the baseline system, and also outperforms other compared domain adaptation techniques. Especially, comparing to the the traditional DAT in~\cite{wang2018unsupervised}, +4.0\% relative error reduction on EER, and a +11.6\% relative improvement on DCF10, as well as a +10.7\% relative improvement on DCF08 are achieved by this solution. The experimental results indicate that when both of multi-source and multi-target domain information is taken into consideration, domain mismatches and data variability can be largely alleviated.
Considering the inner variances of each datasets makes the domain adaptation more efficient and can help to project the samples of all the domains into a more common space.

\vspace{-8pt}
\begin{table}
\centering
\caption{Proposed method vs. other unsupervised domain adaptation methods}
\vspace{-8pt}
\label{table:mdat}
\begin{tabular}{c|c|c|c} 
\toprule
 \textbf{Adaptation Methods}                                                 & \textbf{EER (\%)} & \textbf{DCF10} & \textbf{DCF08}  \\ 
\hline\hline
Interpolated~\cite{garcia2014unsupervised}~\cite{Shon2017Autoencoder}                                                                 & 6.55              & 0.652          & 0.454           \\
IDV~\cite{Kanagasundaram2015Improving}~\cite{Shon2017Autoencoder}                                                                          & 6.15              & 0.676          & 0.476           \\
DICN~\cite{Rahman2015Dataset}~\cite{Shon2017Autoencoder}                                                                         & 4.99              & 0.623          & 0.416           \\
DAE~\cite{Kudashev2016A}~\cite{Shon2017Autoencoder}                                                                          & 4.81              & 0.610          & 0.398           \\
AEDA~\cite{Shon2017Autoencoder}                                                                         & 4.50              & 0.589          & 0.362           \\ 
\hline
DAT~\cite{wang2018unsupervised}                                                                          & 3.73              & 0.541          & 0.335           \\ 
\hline
\begin{tabular}[c]{@{}c@{}}\textbf{MS-DAT}\\(based on LDC code)\end{tabular} & 3.62              & 0.516          & 0.319           \\
\begin{tabular}[c]{@{}c@{}}\textbf{MT-DAT}\\(based on LDC code)\end{tabular} & 3.78              & 0.489          & 0.300           \\
\begin{tabular}[c]{@{}c@{}}\textbf{MDAT}\\(based on LDC code)\end{tabular} & 3.59              & 0.478          & 0.299           \\
\begin{tabular}[c]{@{}c@{}}\textbf{MS-DAT}\\(based on k-means)\end{tabular} & 3.65              & 0.543          & 0.321           \\
\begin{tabular}[c]{@{}c@{}}\textbf{MT-DAT}\\(based on k-means)\end{tabular} & 3.69              & 0.538          & 0.319           \\
\begin{tabular}[c]{@{}c@{}}\textbf{MDAT}\\(based on k-means)\end{tabular} & 3.58              & 0.525          & 0.301           \\
\bottomrule
\end{tabular}
\vspace{-12pt}
\end{table}
%\vspace{-8pt}

%\subsection{The effect of $\lambda$ in MAUDA}
%\label{ssec:subhead}

%We also investigate the impact of the hyper-parameters $\lambda$, which used to trade off the two losses, on the performance of MADA (System 13). We select different values of $\lambda$ and dynamic $\lambda$. The impact of $\lambda$ on EER, DCF10 as well as DCF08 are depicted as Fig.~\ref{fig:EER} and Fig.~\ref{fig:DCF}. When $\lambda$=0, the domain predictor is not trained. We can see the EER is decreasing when the increase of the $\lambda$ until we get the lowest EER when $\lambda$=0.5, and the DCF10 and DCF08 is decreasing when the increase of the $\lambda$ until we get the lowest when $\lambda$=0.4. And we also show how the dynamic $\lambda$ will affect the result. 
%\vspace{-12pt}

%\subsection{Distribution of source and target data using domain adversarial training}

\section{Conclusions}
In this study, we proposed adversarial training for multi-domain speaker recognition to eliminate multi-domain mismatches among different subsets in speaker recognition. Compared to previous domain adaptation studies, we take inner dataset variance into consideration and extract multi-domain-invariant and speaker-discriminative representations for the speaker recognition. 
With the proposed methods, we obtained +36.7\% and +4.0\% relative EER improvement compared to baseline system and conventional DAT method respectively. Moreover, our approach improved DCF10 and DCF08 yields up to +11.6\% and +10.7\% compared to DAT. These results suggest the effectiveness of the multi-domain-invariant and speaker-discriminative speech representations in speaker recognition.
%Experimental results on DAC13 dataset indicate that the proposed method is not only effective to solve the multiple domain mismatch problem, but also outperforms the compared unsupervised domain adaptation methods.
In our future work, we will combine our method with teacher-student learning and conduct experiments on more complex scenarios.
%\section{Acknowledgements}

\bibliographystyle{IEEEtran}

\bibliography{mybib}

% \begin{thebibliography}{9}
% \bibitem[1]{Davis80-COP}
%   S.\ B.\ Davis and P.\ Mermelstein,
%   ``Comparison of parametric representation for monosyllabic word recognition in continuously spoken sentences,''
%   \textit{IEEE Transactions on Acoustics, Speech and Signal Processing}, vol.~28, no.~4, pp.~357--366, 1980.
% \bibitem[2]{Rabiner89-ATO}
%   L.\ R.\ Rabiner,
%   ``A tutorial on hidden Markov models and selected applications in speech recognition,''
%   \textit{Proceedings of the IEEE}, vol.~77, no.~2, pp.~257-286, 1989.
% \bibitem[3]{Hastie09-TEO}
%   T.\ Hastie, R.\ Tibshirani, and J.\ Friedman,
%   \textit{The Elements of Statistical Learning -- Data Mining, Inference, and Prediction}.
%   New York: Springer, 2009.
% \bibitem[4]{YourName17-XXX}
%   F.\ Lastname1, F.\ Lastname2, and F.\ Lastname3,
%   ``Title of your INTERSPEECH 2020 publication,''
%   in \textit{Interspeech 2020 -- 20\textsuperscript{th} Annual Conference of the International Speech Communication Association, September 15-19, Graz, Austria, Proceedings, Proceedings}, 2020, pp.~100--104.
% \end{thebibliography}

\end{document}